\documentclass{PoS}

\usepackage{lineno}
\usepackage[htt]{hyphenat}
\usepackage{subfigure}

\title{The photon PDF determination within the xFitter framework}

\ShortTitle{The photon PDF deterination within the xFitter framework}

\author{\speaker{Francesco Giuli}\thanks{on behalf of the analysis team}\\
        University of Oxford\\
        E-mail: \email{francesco.giuli@physics.ox.ac.uk}}

\abstract{The {\tt{xFitter}} project (former {\tt{HERAFitter}} project) is an open-source package that provides a framework for the determination of the parton distribution functions (PDFs) of the proton for many different kinds of analyses in Quantum Chromodynamics (QCD). {\tt{xFitter}} version 2.0.0 has recently been released, and offers an expanded set of tools and options. It incorporates experimental data from a wide range of experiments including fixed-target, Tevatron, HERA, and LHC. {\tt{xFitter}} can analyze this data up to next-to-next-to-leading-order (NNLO) in perturbation theory with a variety of theoretical calculations including numerous methodological options for carrying out PDF fits and plotting tools which help visualise the results. In this contribution, a determination of the photon PDF from fits to recent ATLAS measurements of high-mass Drell-Yan dilepton production at $\sqrt{s}$ = 8 TeV using this framework is presented.}

\FullConference{EPS-HEP 2017, European Physical Society conference on High Energy Physics\\
		5-12 July 2017\\
		Venice, Italy}

\begin{document}

\paragraph{xFitter overview}
The essential components that allow us to make theoretical predictions for experimental measurements of protons and hadrons are the Parton Distribution Functions (PDFs). The precision of the PDF analysis has advanced tremendously in recent years, and these studies are now performed with very high precision at NLO and NNLO in perturbation theory. The {\tt{xFitter}} project~\cite{Abramowicz:2015mha} is an open source QCD fit framework that can generate PDF fits, compare existing PDF sets, assess the impact of new data and perform a variety of other tasks. The {\tt{xFitter}} framework has already been used for more than 40 analyses including many LHC studies. {\tt{xFitter}} is continually being updated, and version 2.0.0 (Frozen Frog) was released in March 2017 with many improvements and new features. Among them, it is important to stress that, for example, {\tt{xFitter}} is able to read and write PDFs in the LHAPDF6 format. {\tt{xFitter}} can also generate comparison plots of data versus theory, with a several options for the definition of the $\chi^{2}$ function and the treatment of experimental uncertainties. Within {\tt{xFitter}} is also possible to study the impact of a particular data set or experiment on the PDFs. {\tt{xFitter}} is able to perform PDF profiling and reweighting studies. Furhtermore, it is possible to perform PDF reweighting in {\tt{xFitter}}; this method allows xFitter to update the probability distribution of a PDF uncertainty set (such as a set of NNPDF replicas) to reflect the influence of new data inputs. For the PDF profiling, {\tt{xFitter}} compares data and MC predictions based on the $\chi^{2}$ minimization, and then constraints the individual PDF eigenvector sets taking into account the data uncertainties. Moreover, given the fact that many PDF analyses are now extended out to NNLO QCD ($\mathcal{O}(\alpha_S^{2})$), the NLO QED effects ($\mathcal{O}(\alpha_{QED})$) also become important; indeed, in {\tt{xFitter}}, the DIS structure functions and PDF evolution (computed with the {\tt{APFEL}} program~\cite{Bertone:2013vaa}) are accurate up to NNLO in QCD and NLO in QED, including the relevant mixed QCD+QED corrections. More precisely, the $\mathcal{O}(\alpha_{S}\alpha_{QED})$ and the corrections  $\mathcal{O}(\alpha_{S}^{2})$ to the DGLAP splitting functions on top of the $\mathcal{O}(\alpha_{QED})$ ones are available, as well as the corrections $\mathcal{O}(\alpha_{QED}\alpha_{S}^{2})$, $\mathcal{O}(\alpha_{S}^{2})$  and $\mathcal{O}(\alpha_{QED}^{2}\alpha_{S})$ to $\beta$ functions.

\paragraph{Data, theory and fit inputs}
In~\cite{Giuli}, the photon content of the proton, $x\gamma(x,Q^2)$, is extracted from a PDF fit to the ATLAS measurements of high-mass Drell-Yan (DY) differential cross sections at $\sqrt{s}=8$ TeV~\cite{Aad:2016zzw}, combined with inclusive DIS cross-section data from HERA~\cite{Abramowicz:2015mha}. Information on the gluon and quark content of the proton is provided by the HERA inclusive data, while  a direct sensitivity to the photon PDF is provided by the high-mass DY data. The ATLAS high-mass DY 8 TeV measurements are presented in terms of both the single-differential (1D) invariant-mass distribution, as well as double\- differential (2D) distributions. The NLO QCD and LO QED cross sections have been supplemented by bin-by-bin $K$-factors in order to achieve NNLO QCD and NLO EW accuracy and they have been obtained from {\tt FEWZ}~\cite{Gavin:2012sy}. They are defined as follows:
\begin{equation}
  \label{eq:kfactor}
  K(m_{ll},|y_{ll}|) \equiv\frac{\rm NNLO\  QCD  + NLO\  EW}{\rm NLO\  QCD + LO\  EW} \, ,
\end{equation}
where the MMHT2014 NNLO~\cite{Harland-Lang:2014zoa} PDF set is used both in the numerator and in the denominator. The explicit form of PDF parametrisation at the starting scale $Q_0^2$ is determined by the technique of saturation of the $\chi^{2}$~\cite{Wilks:1938dza} and it has been found that the optimal parametrisation for the gluon and quark PDFs is 
$xu_v(x) = A_{u_v}x^{B_{u_v}}(1-x)^{C_{u_v}}(1+E_{u_v}x^{2}), xd_v(x) = A_{d_v}x^{B_{d_v}}(1-x)^{C_{d_v}}, x\bar{U}(x) = A_{\bar{U}}x^{B_{\bar{U}}}(1-x)^{C_{\bar{U}}}, x\bar{D}(x) = A_{\bar{D}}x^{B_{\bar{D}}}(1-x)^{C_{\bar{D}}}, xg(x) = A_{g}x^{B_{g}}(1-x)^{C_{g}}(1+\\E_{g}x^{2})$. For the photon PDF, the parametrisation $x\gamma(x) = A_{\gamma}x^{B_{\gamma}}(1-x)^{C_{\gamma}}(1+D_{\gamma}x+E_{\gamma}x^{2})$ has been used.
PDF uncertainties are estimated using the Monte Carlo (MC) replica method~\cite{DelDebbio:2004xtd,DelDebbio:2007ee,Ball:2008by}, and it has been cross-checked that the Hessian method~\cite{Pumplin:2001ct} provides consistent results.

\begin{figure}[t]
\centering
\subfigure[\label{MChesse}]{\includegraphics[width=7.5cm]{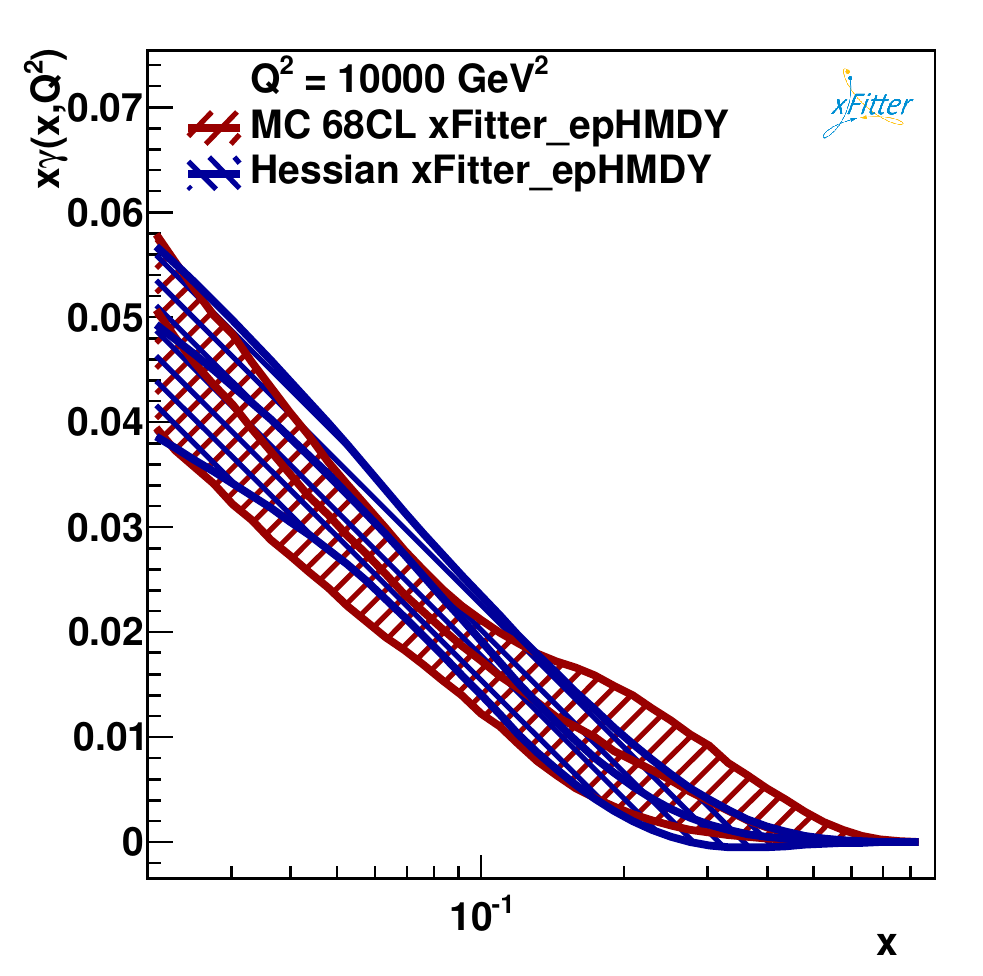}}
\subfigure[\label{nlo_nnlo}]{\includegraphics[width=7.5cm]{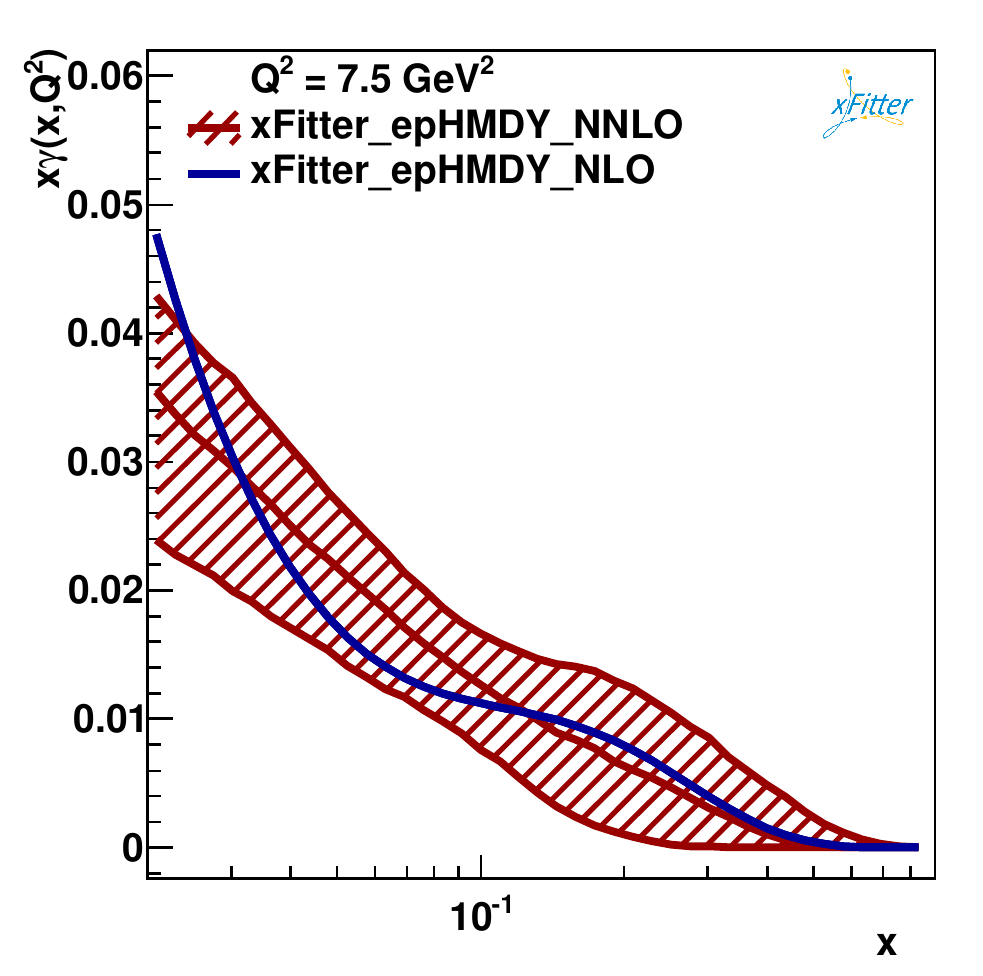}}
\caption{(a) Comparison between the {\tt{xFitter\_epHMDY}} determinations at $Q^2=10^{4}$ GeV$^2$, obtained with the MC (baseline) and with the Hessian methods, where the PDFerror band shown corresponds to the 68\% C.L. uncertainties  in both cases; (b) Comparison between the reference {\tt{xFitter\_epHMDY}} fit of $x\gamma(x,Q^2)$, based on  NNLO QCD and NLO QED theoretical calculations, and the central value of the corresponding fit based on NLO QCD and QED theory, at $Q^2=7.5$ GeV$^2$. In the former case, only the experimental MC PDF uncertainties are shown.}
\end{figure}

\paragraph{Results}
\begin{figure}[t]
\centering
\includegraphics[width=7.5cm]{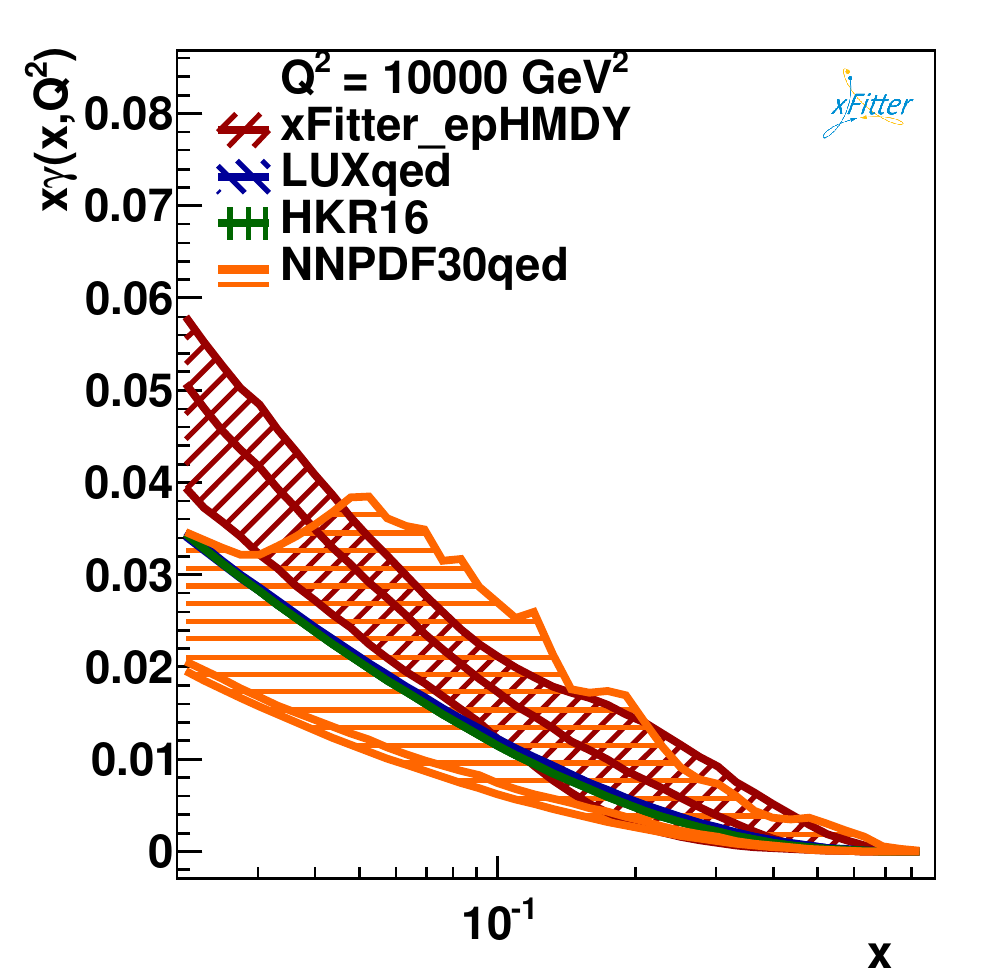}
\includegraphics[width=7.5cm]{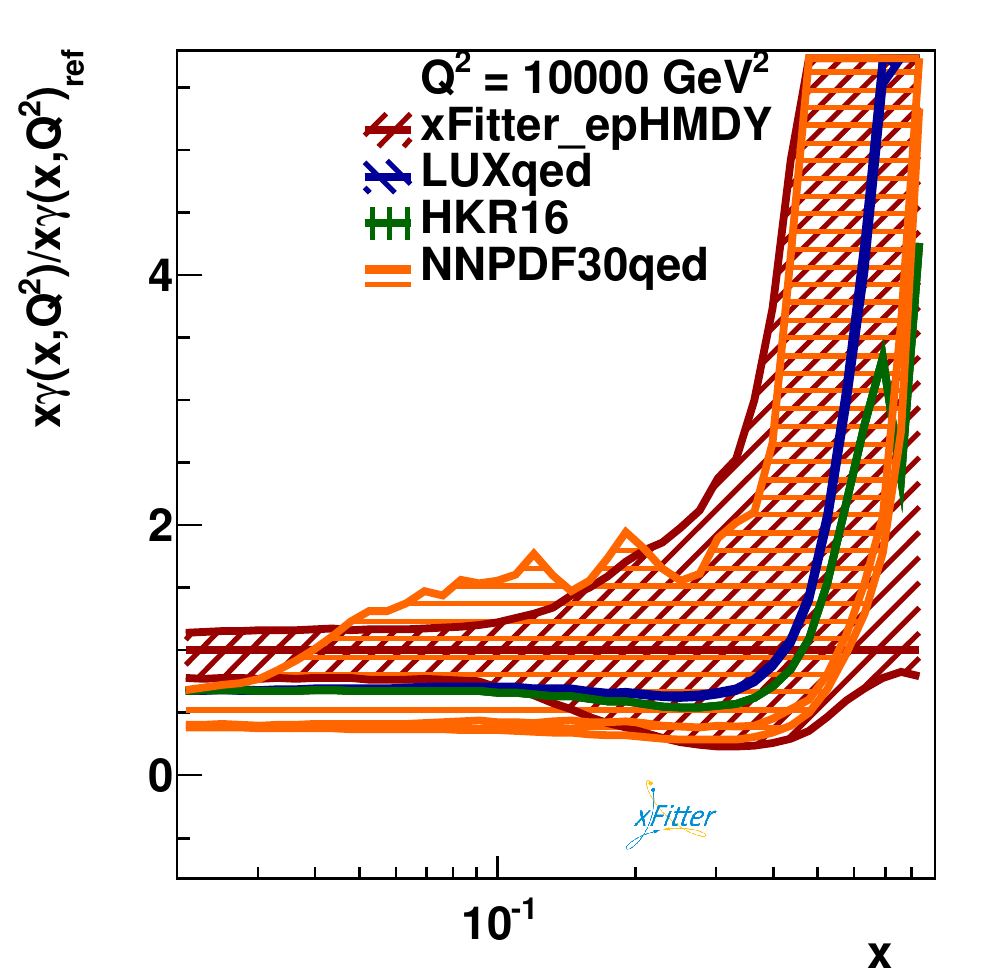} 
\caption{Left: Comparison between the photon $x\gamma(x,Q^2)$ at $Q^2=10^4$ GeV$^2$ from the present NNLO analysis ({\tt xFitter\_epHMDY}) with the corresponding results from NNPDF3.0QED, LUXqed and HKR16. Right: the same comparison, now with the results normalized to the central value of {\tt xFitter\_epHMDY}. For the present fit, the PDF uncertainties are shown at the 68\% CL obtained from the MC method. For HKR16 only the central value is shown, while for LUXqed the associated PDF uncertainty band is included. }
\label{photon_zoom} \label{photon_zoom_ratio}
\end{figure}
In the following, the results that will be shown correspond to those obtained from fitting the double-differential $(m_{ll},y_{ll})$ cross-section distributions. The value of $\chi^2_{min}/N_{dof} = 1284/1083$ has been obtained for the baseline NNLO fit where $N_{dof}$ represnts the number of degrees of freedom. In Fig.~\ref{photon_zoom}, the photon PDF, $x\gamma(x,Q^2)$, is shown at the evolved scale $Q^2=10^4$ GeV$^2$,  and it is compared to the predictions obtained using HKR16~\cite{hkr2016}, LUXqed~\cite{Manohar:2016nzj,Manohar:2017eqh} and NNPDF3.0QED~\cite{Ball:2013hta}. The {\tt xFitter\_epHMDY} fit is shown with the associated experimental PDF uncertainties at the 68\% confidence level (CL), obtained using the MC replica method. For limited sensitivity to the photon PDF, the $x$-range is set between 0.02 and 0.9 in Fig.~\ref{photon_zoom}. The four determinations of the photon PDF shown in Fig.~\ref{photon_zoom} are consistent within PDF uncertainties for $x\ge 0.1$; for smaller values, the baseline NNLO fit is in agreement at the 2-$\sigma$ level with the LUXqed and HKR16 predictions. It has been shown in fig.~\ref{photon_zoom} that for $x$ between $0.04$ and $0.2$ the present analysis exhibits smaller PDF uncertainties as compared to those predicted by NNPDF3.0QED.\\
In order to test the robustness of this photon PDF determination, a number of variations has been assessed. The $x\gamma(x, Q^2)$ determination is compared with further fits, where a number of new parameters are allowed in the PDF parametrisation. In each case, one variation at a time is performed and compared with the central value of $x\gamma(x,Q^2)$ and its experimental PDF uncertainties computed using the MC method. The result of these studies show that all the different variations are contained within the experimental PDF uncertainty bands. Then, the central fit prediction has been compared with the central value of those fits for which the theory input parameters (the input parametrisation scale $Q_0^2$, the strong coupling constant $\alpha_{S}$, the values of the $m_{c}$ and $m_{b}$, the ratio of strange to non-strange light quark PDFs and the minimum cut on $Q^2$ of data to be entered in the fit) have been varied. It has been found that the effect of the variations considered is contained within the experimental PDF uncertainty bands of the reference fit. Moreover, a comparison between the MC and Hessian methods is provided, in order to estimate the robustness of the estimated experimental uncertainty of $x\gamma(x, Q^2)$ in this analysis. Fig.~\ref{MChesse} shows this comparison at the evolved scale of $Q^2=10^{4}$ GeV$^2$ and a reasonable agreement between the two methods has been indicated (it has been cross-checked that a similar results can be found at the lower scale $Q^2=7.5$ GeV$^2$). As expected, the MC uncertainties are larger than the ones computed with the Hessian method, indicating deviations with respect to the Gaussian behaviour of the photon PDF. In the end, the perturbative stability of the {\tt xFitter\_epHMDY} photon PDF fit with respect to the inclusion of higher order QCD corrections has been quantified. A comparison between the baseline fit of $x\gamma(x,Q^2)$, based on NNLO QCD and NLO QED theoretical calculations, with the central value resulting from a corresponding fit based instead on NLO QCD and QED theory has been made. A reasonable perturbative stability is exhibited by the {\tt xFitter\_epHMDY} photon PDF determination; indeed, the central value of the NLO fit is always contained in the 1-$\sigma$ PDF uncertainty band of the baseline {\tt xFitter\_epHMDY} fit. The agreement between the two fits is particularly good for $0.1\lesssim x$. Fig.~\ref{nlo_nnlo} shows this comparison at the starting scale, $Q^2=7.5$ GeV$^2$; a similar result has been found at the higher scale, $Q^2=10^{4}$ GeV$^2$, indicating that perturbative stability is not scale dependent.

\paragraph{Acknowledgements}
I would like to thank Voica Radescu and Ringaile Placakyte for their contribution to this work and for their dedication as conveners of the {\tt{xFitter}} project from 2012 until May 2017.

\end{document}